\documentstyle[manuscript,aps]{revtex}
\begin{document}

\title{On the Injection Energy Distribution of Ultra-High-Energy Cosmic
Rays}
\author{James Geddes, Theodore C. Quinn, and Robert M. Wald}
\address{Enrico Fermi Institute and Department of Physics\\
         University of Chicago\\
         5640 S. Ellis Avenue\\
         Chicago, Illinois 60637-1433}
\date{}

\maketitle
\begin{abstract}
We investigate the injection spectrum of ultra-high-energy ($>10^{15}
\,{\rm eV}$) cosmic rays under the hypotheses that (1)~these cosmic
rays are protons and (2)~the sources of these cosmic rays are
extra-galactic and are homogeneously distributed in space, although
they may have had a different strength in the past; furthermore, we
assume that we are not unusually close to any individual
source(s). The most puzzling aspect of the observed ultra-high-energy
cosmic ray spectrum is the apparent nonexistence of a ``Greisen
cut-off'' at about $10^{19.8} \,{\rm eV}$. Such a cut-off would be
expected due to rapid energy loss from photopion production caused by
interactions with the microwave background. We show that this fact
could be naturally explained if most (or all) of the cosmic rays
presently observed above about $10^{19.6} \,{\rm eV}$ were initially
injected with energy above the Greisen cut-off. However, we find that
the injection of cosmic rays above the Greisen cut-off cannot account
for the observed flux below about $10^{19.6} \,{\rm eV}$ unless the
injection rate of these particles was enormously higher in the past,
as would be the case if the injection resulted from the decay of an
ultra-massive particle with lifetime of order $10^{9} \,{\rm
yr}$. Even with such a rapid source evolution, the observed cosmic ray
spectrum below about $10^{18.5} \,{\rm eV}$ cannot be explained by
injection of particles above the Greisen cut-off in the distant
past. However, we show that a $1/E^3$ injection spectrum can account
for the observed spectrum below $10^{18.5} \,{\rm eV}$, with the
steepening observed by the Fly's Eye group between $10^{17.6}\rm\,eV$
and~$10^{18.5}\rm\,eV$ being very naturally explained by $e^+e^-$
production effects. This latter fact lends support to the hypothesis
that the cosmic rays in this energy regime are protons.  However, due
to $e^+e^-$ production effects, a $1/E^3$ injection spectrum cannot
account for the observed flux above about $10^{18.5} \,{\rm eV}$.
\end{abstract}

\newpage

\section{Introduction}
The existence of ultra-high-energy cosmic ray particles is certainly
one of the most remarkable phenomena observed in nature. Their
existence at energies above about $10^{19.8} \,{\rm eV}$\cite{fe2,ake}
is particularly mysterious since, if the cosmic rays are protons,
photopion production caused by interactions with the microwave
background should result in a rapid loss of energy and consequent
depletion of the observed flux of these particles
\cite{gre,zk}. Similar serious difficulties in accounting for the
existence of such cosmic rays occur under the various alternative
hypotheses concerning the nature of the cosmic ray particles; see,
e.g., the discussions in \cite{ssb} and \cite{es}.

It seems clear that the explanation of the existence of the highest
energy cosmic rays will require some unconventional ideas, or, at
least, the extrapolation of conventional ideas to unconventional
extremes. In the absence of a reliable theoretical framework, it is
difficult to make arguments concerning the plausibility of various
hypotheses. Furthermore, the experimental data on the highest energy
cosmic rays suffers from poor statistics as well as from significant
uncertainties in energy determinations, so there are very few
``facts'' that can be pinned down with complete certainty.

Nevertheless, under suitable hypotheses about the nature of the cosmic
ray particles and the distribution of their sources in space and time,
it is possible, in principle, to say a great deal about the energy
distribution which the cosmic rays must have possessed at the time
they were injected into the universe, since the energy loss rate of
cosmic rays is governed by well established physics. In this paper, we
shall investigate the injection energy distribution of cosmic rays of
present energy $>10^{15} \,{\rm eV}$ under the hypotheses that
\begin{enumerate}
\item these cosmic rays are protons \footnote{It should be noted,
however, that some (model dependent) evidence was reported in
\cite{fe1} that the composition of the the cosmic rays below
$10^{18.5} \,{\rm eV}$ tends toward heavy nuclei.}
\item the sources of these cosmic rays are extra-galactic and are
homogeneously distributed in space. Furthermore, we are not unusually
close to any individual source(s) so that the cosmic ray flux we see
is representative of that occurring elsewhere in the universe.
\end{enumerate}
In our analysis we allow for the possibility that the sources may have
been more numerous (or less numerous) in the past but we assume that
the shape of the injection energy spectrum of the sources does not
vary with time. However, except for the decaying particle model
considered at the end of Section~\ref{HighEnergyResults}, source
evolution will not play an important role in any of our
arguments. Note that our assumptions automatically give rise to a
homogeneous, isotropic cosmic ray population. Consequently, under our
hypotheses, the possible presence of an intergalactic magnetic field
will have no effect upon the cosmic ray spectrum.

Even under these assumptions we cannot, in principle, uniquely
determine the injection spectrum, since the presently observed
spectrum depends upon both the injection spectrum and the time history
of the sources. Furthermore, as already mentioned above, the detailed
structure of the observed spectrum at the highest energies is quite
uncertain, so even if a mathematical inversion could be done, it
probably would not have much significance. Nevertheless, we shall see
that some quite general, nontrivial constraints on the injection
spectrum and source evolution can be obtained.

Under our two assumptions above, the following two things can be
concluded immediately about the cosmic ray injection spectrum: (a) In
the energy range from about $10^{15} \,{\rm eV}$ to about $10^{17.6}
\,{\rm eV}$, the observed spectrum is well fit by a $1/E^3$ power
law. In this energy range, the only significant energy loss mechanism
for protons is cosmological redshift. Since redshift takes a power law
spectrum (injected at any initial time) to a power law of the same
power, we can conclude that in this ``low energy'' regime, the injection
spectrum of the protons follows a $1/E^3$ power law.  (b) Above about
$10^{19.8} \,{\rm eV}$ in the present universe, protons rapidly lose
energy due to photopion production reactions with the microwave
background. Nevertheless, cosmic rays are observed above this
``Greisen cut-off''. Indeed, the observations of two events well above
$10^{20}\rm\,eV$ have been reported recently \cite{fe2,ake}.
Thus, assuming, as above, that these cosmic rays are protons, we can
conclude that, in the present universe, protons are being injected
with energies above the Greisen cut-off at a rate that is directly
calculable from the observed flux and the photopion production energy
loss rate (see Section~\ref{HighEnergyResults}).

The injection rate of above-Greisen-cut-off protons obtained from (b)
lies far above extrapolations of the $1/E^3$ injection spectrum
deduced from (a). Thus, there is no reason to expect that the ``low
energy'' (i.e., between $10^{15} \,{\rm eV}$ and $10^{17.6} \,{\rm
eV}$) and highest energy (i.e., $>10^{19.8} \,{\rm eV}$) cosmic rays
are produced by a common mechanism. Hence, we shall presume that there
are (at least) the following two independent sources of cosmic rays
(though we do not exclude the possibility that these sources
correspond to the same physical objects or phenomena): (i) A ``low
energy''source which injects protons with a $1/E^3$ power law. (ii) A
``high energy'' source which injects protons with energies above the
Greisen cut-off. Our main goal in this paper is to investigate how far
one can go toward explaining the entire high energy cosmic ray
spectrum by (simple extrapolations of) these two sources.

Our main results are the following. First, in
Section~\ref{HighEnergyResults}, we investigate the contributions of
the high energy source.  Specifically, we compute the cosmic ray
spectrum which would be observed if protons with energies above the
Greisen cut-off are injected at a rate corresponding to the presently
observed above-Greisen-cut-off flux, and if this rate does not vary
with time.  We find that the predicted spectrum is quite compatible
with the observed spectrum at energies down to about $10^{19.6} \,{\rm
eV}$. This implies that, under our two assumptions above, many---and,
quite possibly, all---of the cosmic rays with energies above
$10^{19.6} \,{\rm eV}$ were initially injected with energy above the
Greisen cut-off. However, the predicted spectrum below $10^{19.6}
\,{\rm eV}$ lies well below the observed spectrum. We cannot
significantly improve the predicted spectrum by assuming that the high
energy source injects protons with a power law spectrum (rather than
injecting all the protons at energies above the Greisen cut-off),
since a hard injection spectrum (say, $1/E^2$ or harder) also gives
poor agreement below $10^{19.6} \,{\rm eV}$, whereas a soft injection
spectrum (say, $1/E^2$ or softer) produces a dramatic Greisen
cut-off. However, we show that good agreement with the observed
spectrum down to an energy of about $10^{18.5} \,{\rm eV}$ would be
obtained if the injection rate of protons per comoving volume was
significantly higher in the past. Indeed, to get good agreement it is
necessary for the high energy source to be about 200 times stronger at
a redshift $z = 1/2$.  This is much too rapid an increase in the
injection rate to plausibly attribute to increased activity of active
galactic nuclei or other possible astrophysical sources of cosmic
rays. However, if the injection of above-Greisen-cut-off protons
results from the decay of an ultra-massive particle, the required
source strengthening would correspond to a lifetime of about $10^9
\,{\rm yr}$. The existence of such a particle with this lifetime and
with the mass and abundance required to produce the highest energy
cosmic rays appears to be compatible with known constraints, though
remarkably close to the limit obtained from the gamma ray background
\cite{ellis}.

Second, in Section~\ref{LowEnergyResults}, we investigate the
contributions of the low energy source to the observed cosmic ray
spectrum.  Specifically, we analyze the extent to which an
extrapolation of the $1/E^3$ injection spectrum to arbitrarily high
energies could account for observed spectrum above $10^{15} \,{\rm
eV}$. We show that the steepening in the spectrum observed by the
Fly's Eye group \cite{fe1} between about $10^{17.6} \,{\rm eV}$ and
$10^{18.5} \,{\rm eV}$ is naturally accounted for by $e^+ e^-$
production effects. However, we find that the cosmic ray flux between
$10^{18.5} \,{\rm eV}$ and $10^{19.6} \,{\rm eV}$ is significantly
depleted by these effects, so a $1/E^3$ (or softer) injection spectrum
cannot explain the observed spectrum above about $10^{18.5} \,{\rm
eV}$.

In Section~\ref{Methods} we describe the methods and approximations
used to calculate the relationship between the injected and observed
cosmic ray spectra. The contribution to the observed cosmic ray
spectrum of protons initially injected with energy above the Greisen
cut-off is analyzed in Section~\ref{HighEnergyResults}, and the
contribution from protons initially injected with a $1/E^3$ spectrum
is analyzed in Section~\ref{LowEnergyResults}. Our conclusions are
summarized in Section~\ref{Conclusions}.

\section{Methods}
\label{Methods}

There are three potentially significant causes of energy loss for
ultra-high energy cosmic ray protons propagating in intergalactic
space: (1) Cosmological redshift, (2)~$e^+e^-$ pair production
scattering with the cosmic microwave background radiation (CMBR), and
(3)~pion production scattering with the CMBR. These effects have been
calculated previously by many
authors\cite{gre,zk,blum,hs,bg,steck,yt,ac} under various
approximations and computational schemes. The purpose of this section
is to describe our calculation of these effects in some detail.

The energy loss due to cosmological redshift is a continuous process,
which is governed by a simple differential equation. The energy loss
due to $e^+e^-$ pair production is a stochastic process, but the pair
production cross-section is high and the fractional energy lost per
scattering event is very small (on the order of~$2m_e/m_p \simeq
10^{-3}$ at threshold and decreasing thereafter).  Hence, a
continuous, mean energy loss approximation should be excellent for
treating the effects of process (2), as we have verified by comparing
the results of our calculations to the Monte Carlo calculations of
Yoshida and Teshima \cite{yt} (see subsection~\ref{Numerics}
below). On the other hand, for process (3), the fraction of energy
lost per pion production event is~$0.13$ at threshold and rises
thereafter\cite{gre}, so, in general, statistical fluctuations are of
importance for calculating the effects of pion production
scattering. However, in this paper, we restrict consideration to
spatially homogeneous and temporally continuous injection, which
provides a smoothing that mimics the smoothing provided by the
statistical fluctuations. Consequently, in our case, the continuous,
mean energy loss approximation\cite{bg} should be adequate for
treating process (3), as we have verified by comparing our results
with a Monte Carlo calculation provided to us by
F. Aharonian (see below). We now describe in more detail
the formulas we used to calculate the energy loss rate of cosmic ray
protons.

\subsection{Energy loss processes}

\subsubsection{Cosmological redshift}

We take our cosmological model to be a standard, matter-dominated,
Robertson-Walker universe, with no cosmological constant.  As we shall
see below, for the source time evolutions we consider, the
contribution of cosmic rays injected at redshift,~$z$, greater than
about~$1/2$ will not be of great importance. Hence, our results should
not be very sensitive to the precise value of the closure parameter,
$\Omega$, and we shall simply set $\Omega = 1$, corresponding to a
spatially flat universe with scale factor $a(t)$ given by
\begin{equation}
\frac{a(t)}{a(t_0)} = \left(\frac{t}{t_0}\right)^{2/3}.
\end{equation}

Our calculations are somewhat sensitive to the assumed value of
Hubble's constant.  For most of our calculations, we assumed that the
universe is 15 billion years old ($t_0=4.7\times10^{17}\,{\rm s}$),
corresponding to a ``low'' value of Hubble's constant ($H_0=43.5\,{\rm
km}\,{\rm s}^{-1}\,{\rm Mpc}^{-1}$). However, we also recalculated
most effects using a Hubble constant of~$75\,{\rm km}\,{\rm
s}^{-1}\,{\rm Mpc}^{-1}$ in order to verify that our conclusions did
not depend sensitively on~$H_0$.

The energy of a relativistic particle propagating through the universe
scales inversely with~$a(t)$, so we have an energy loss rate due to
cosmological redshift given by
\begin{equation}
\left(\frac{dE_p}{dt}\right)_{\rm rs} = -\frac{2E}{3t},
\end{equation}
In particular, it should be noted that the attenuation length for
cosmological redshift is independent of energy.

\subsubsection{Pair production scattering}

Calculations of proton energy loss due to pair production scattering
with the CMBR have been performed by Blumenthal\cite{blum}, by
Berezinsky and Grigor'eva\cite{bg}, and others. We were unaware of
this prior work when we began our investigations, and derived our mean
energy loss formula independently. Since our formula differs slightly
from that of other authors, we present the derivation of it
here.

In the ``lab frame'' (i.e., the isotropy frame of the CMBR), the
photons of the CMBR have a number density per energy per solid angle
per volume given by the standard Planck formula
\begin{equation}
n(E)\,dE\,\sin\theta\,d\theta\,d\phi\,dV=
	\frac{1}{(2\pi\hbar c)^3}\frac{2E^2}{e^{E/kT} - 1}
	\,dE\,\sin\theta\,d\theta\,d\phi\,dV
\end{equation}
In the rest frame of a proton travelling with velocity
parameter~$\beta$ in the~$+z$ direction, this distribution corresponds
to
\begin{equation}
\bar{n}(\bar{E},\bar{\theta})
\,d\bar{E}\,\sin\bar{\theta}\,d\bar{\theta}\,d\bar{\phi}\,d\bar{V} =
\frac{1}{(2\pi\hbar c)^3}
\frac{2\bar{E}^2}{\exp\left[\frac{\gamma\bar{E}}{kT}
                   (1+\beta\cos\bar{\theta})\right] - 1}
\,d\bar{E}\,\sin\bar{\theta}\,d\bar{\theta}\,d\bar{\phi}\,d\bar{V}.
\end{equation}
where the bars denote the corresponding variables in the proton
frame. Since we are only interested in ultra-relativistic protons,
this distribution will be very sharply peaked about~$\bar{\theta} =
\pi$, i.e., essentially all of the blackbody photons will be incident
head-on in the proton frame. Integrating over the angles, we
obtain~$f(\bar{E})$, a number density per energy per volume in the
proton frame,
\begin{equation}
f(\bar{E})\,d\bar{E}\,d\bar{V}= \frac{4\pi}{(2\pi\hbar c)^3}
         (kT)^2\frac{\xi}{\beta}
         \ln\left(
         \frac{1 - \exp[-\xi(1-\beta)^{-1}]}
              {1 - \exp[-\xi(1+\beta)^{-1}]}
         \right)\,d\bar{E}\,d\bar{V},
\end{equation}
where~$\xi=\bar{E}/\gamma kT$. Finally, we take the ultra-relativistic
limit ($\beta \rightarrow 1$), giving us
\begin{equation}
f(\bar{E}) = \frac{4\pi}{(2\pi\hbar c)^3}
       (kT)^2 \xi
       \ln\left(
       \frac{1}
            {1 - e^{-\xi/2}}
       \right).
\label{bb}
\end{equation}

We now drop the bars on the proton frame variables and denote the
energy of an incident photon as~$E_{\gamma}$. In the limits
that~$\beta \approx 1$ but~$E_{\gamma} \ll m_p c^2$, the energy loss
of the proton in the lab frame due to a single scattering event is
given by
\begin{equation}
E_{\rm loss} = \gamma p_p c \cos\theta,
\end{equation}
where~$p_p$ is the recoil momentum of the proton in the proton frame
and~$\theta$ is the recoil angle. The probability distributions of
these recoil variables as a function of~$E_{\gamma}$ were provided by a
second order QED calculation given by Jost et al.\cite{jost}.  They
obtained the following expression for the differential cross-section
as a function of the recoil variables (see their Eq.~(47))
\begin{eqnarray}
\frac{d\sigma}{dQd\eta} &=&
\frac{\alpha r_{\rm e}^2}{\epsilon^2}
\frac{1}{Q^2}
\left\{
  \ln\frac{1-\omega}{1+\omega}
  \left[
    \left(1-\frac{\epsilon^2}{\eta^2}\right)
    \left[
      1-\frac{1}{4\eta^2}+\frac{1}{2\eta Q}-\frac{1}{8Q^2\eta^2}
      -\frac{Q}{\eta}+\frac{Q^2}{2\eta^2}
    \right] + \frac{\epsilon^2}{2\eta^4}
  \right] \right.  \\
\nonumber
&&\left.\phantom{\frac{1}{Q^2}\frac{\alpha r_{\rm e}^2}{\epsilon^2}\{}
  + \omega
  \left[
    \left(1-\frac{\epsilon^2}{\eta^2}\right)
    \left(1-\frac{1}{4\eta^2}+\frac{1}{2\eta Q}\right)
    +\frac{1}{\eta^2}
    \left(1-\frac{2\epsilon^2}{\eta^2}\right)
    (-2Q\eta+Q^2)
  \right]
\right\}
\end{eqnarray}
where~$\epsilon = E_{\gamma}/2m_ec^2$,~$Q=p_p/2m_ec$, $\eta = \epsilon
\cos\theta$, and~$\omega = [1 - 1/(2Q\eta - Q^2)]^{\frac{1}{2}}$. In
terms of these~$\epsilon$,~$Q$, and~$\eta$ variables, our single event
lab frame energy loss is given by
\begin{equation}
E_{\rm loss} =\frac{2 m_e c^2\gamma}{\epsilon}Q\eta.
\end{equation}
Therefore, at a given photon energy~$\epsilon$ we obtain a mean energy
loss by integrating~$E_{\rm loss}$ against the differential cross
section and then dividing by total cross section.
\begin{equation}
\langle E_{\rm loss} \rangle
 =\frac{1}{\sigma}\frac{2 m_e c^2\gamma}{\epsilon}
    \int_1^\epsilon \!d\eta
    \int_{\eta-\sqrt{\eta^2-1}}^{\eta+\sqrt{\eta^2-1}} \!dQ
    \,\left(Q\eta\frac{d\sigma}{dQd\eta}\right)
\end{equation}
where the limits of the integrals are set by kinematic constraints.

Finally, to calculate~$(dE_p/dt)_{e^+e^-}$, we multiply our number
density~$f(E_{\gamma})$ by~$c$ to produce a flux of photons per photon
energy and integrate against~$\langle E_{\rm loss} \rangle \sigma$,
remembering to insert a factor of~$1/\gamma$ to convert from an event
rate in the proton frame to an event rate in the lab frame:
\begin{equation}
\left(\frac{dE_p}{dt}\right)_{e^+e^-} = -\frac{c}{\gamma}
                \int_{2 m_e c^2}^{\infty}
                \langle E_{\rm loss} \rangle\,\,
                \sigma\,\,
                f\,\,
                dE_{\gamma}.
\end{equation}

Our energy loss formula differs slightly from that of
Blumenthal\cite{blum} in that we used the exact expression (\ref{bb})
for a highly blueshifted black body distribution, and it also differs
slightly from that of Berezinsky and Grigor'eva\cite{bg} in that they
made some approximations to the formula for the
cross-section. However, we obtained excellent agreement with the
results of both Blumenthal and Berezinsky and Grigor'eva when we
evaluated the energy loss rate numerically for a range of proton
energies at CMBR temperature~$T=T_0=2.73\,{\rm K}$.

\subsubsection{Pion production scattering}

A formula for the energy loss due to photopion production can be
obtained in parallel with our above derivation for the energy loss due
to $e^+e^-$ production. However, analytic expressions for the
differential cross section as a function of recoil momentum are
unavailable for pion production scattering, so one is forced to rely
on experimental data. In our calculations, we used the approximations
due to Berezinsky and Grigor'eva\cite{bg}, which make use of the fact
that, at the proton energies of relevance here, the photopion
production predominantly occurs near threshold. Hence, it should be a
reasonable approximation to treat the photopion production as being
isotropic in the rest frame of the proton, and to treat the total
cross-section as a linearly rising function of energy, i.e., $\sigma
(E) = (E - E_{th})\sigma'$.  This yields the following simple formula
for the energy loss rate,~$(dE_p/dt)_{\pi}$, due to photopion production
in terms of the single parameter $\sigma'$ (cf.\ Eq. (8) of
\cite{bg})
\begin{equation}
\left(\frac{dE_p}{dt}\right)_{\pi} =
		\frac{2 (kT)^3 \sigma' \epsilon_0^2 \gamma}{\pi^2 c^4 \hbar^3}
                \exp(-\epsilon_0/2\gamma kT)
\end{equation}
where~$\epsilon_0 = m_{\pi}(1+m_{\pi}/m_p)$.  Berezinsky and
Grigor'eva used~$\sigma' = 6.8\times10^{-36}\,{\rm cm}^2/{\rm eV}$ for
their calculations, but we used~$\sigma' = 3.45\times10^{-36}\,{\rm
cm}^2/{\rm eV}$ in order to more closely approximate the cross
section data given by Hikasa et al.\cite{pioncsdata}.

\begin{figure}
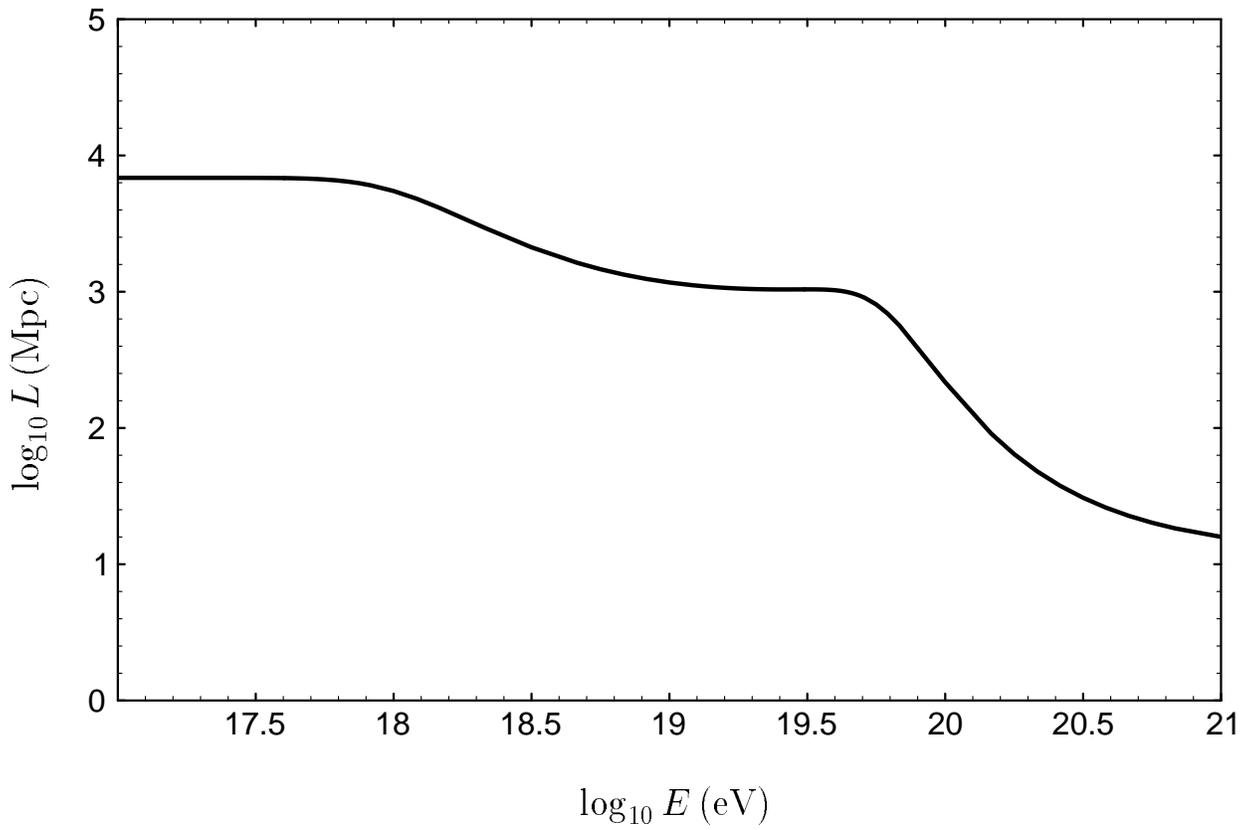

\caption{The attenuation length for a proton travelling through
intergalactic space in the present universe. The energy loss is due to
pion production scattering, pair production scattering, and
cosmological redshift.}
\label{AttLength}
\end{figure}

As a check on the validity of this approximation we plotted the
attenuation length as a function of proton energy at~$T=T_0$ through
the range where pion production dominates (see Figure~\ref{AttLength})
and obtained excellent agreement with the Monte Carlo results of
Yoshida and Teshima\cite{yt} up to a proton energy of~$10^{21}\,\rm
eV$.

\subsection{Calculation of the present spectrum of cosmic rays}

The present energy spectrum of cosmic ray protons is determined from
the injection rate of protons per comoving volume per energy per
time,~$I(t,E)$, by integrating the mean energy loss equation
\begin{equation}
\frac{dE_p}{dt} =  \left(\frac{dE_p}{dt}\right)_{\rm rs}
		   + \left(\frac{dE_p}{dt}\right)_{e^+e^-}
                   + \left(\frac{dE_p}{dt}\right)_{\pi},
\label{mel}
\end{equation}
where~$E_p(t)$ is the energy of the proton. The expressions for the
various terms on the right side were given in the previous subsection.
The information contained in this equation is most usefully encoded by
expressing the initial energy,~$E'$, in terms of the energy
today,~$E$, and the injection time,~$t$; that is, by
finding~$E'(E,t)$. This we did by numerically solving~(\ref{mel}) with
the boundary condition $E_p(t_0) = E$.

The calculations of this paper assume that the sources of cosmic rays
are homogeneously distributed throughout the universe. In addition, we
assume that $I(t,E)$ is of the form $I(t,E) = h(t) g(E)$, so that the
only time dependence in the injection spectrum is the overall rate of
injection. (This would be the case if the nature of the sources of the
cosmic-rays did not change but the number of such sources did.) By
definition, the total number of protons per comoving volume
element,~$dn$, injected between times~$t$ and~$t+dt$ with energies
between~$E'$ and~$E'+dE'$ is
\begin{equation}
dn(E',t)  = h(t) g(E') \,dt \,dE'.
\end{equation}
Hence, the present number density spectrum of cosmic rays arising from
those injected between times~$t$ and~$t+dt$ is given by
\begin{equation}
dn(E,t) = h(t) g(E')
	\frac{\partial E'}{\partial E}\,dE\,dt,
\end{equation}
where $E'(E,t)$ is obtained from integration of (\ref{mel}) as
described above.  This expression can then be integrated over all
times to give the comoving number density of protons today. Since
these protons are distributed homogeneously and isotropically and are
all travelling at approximately~$c$, the differential flux
per energy per time per steradian today,~$J(E)$, is given by
\begin{equation}
J(E)	= \frac{c}{4\pi} \frac{dn}{dE}
	= \frac{c}{4\pi} \int^{t_0} h(t) g(E')
	\frac{\partial E'}{\partial E}\,dt.
\label{fluxintegral}
\end{equation}
\subsection{Numerical methods}
\label{Numerics}

All of our numerical calculations were performed using Mathematica on
a Silicon Graphics workstation. In particular, the differential
equation (\ref{mel}) was integrated using Mathematica's NDSolve
routine for a large number of injection energies and times to produce
a discrete version of the map~$E'(E,t)$. The contribution from each
time was computed by calculating the derivative~$\partial E'/\partial
E$ point by point, and then the integral (\ref{fluxintegral}) was
approximated by simply adding these contributions multiplied by the
appropriate~$\Delta t$ values.

As a check on all of the above, as well as our energy loss formula, we
set the parameters of our injection function~$I(E,t)$ to match those
considered by Yoshida and Teshima in their Monte Carlo
calculations\cite{yt} and attempted to reproduce curves 1,2 and 4 of
their Figure~5 (For this calculation we used their value
of~$H_0=75\rm\,km\,s^{-1}\,Mpc^{-1}$).  Agreement in all three cases
was generally better than 10\% at all energies, although somewhat
larger discrepancies occurred at energies corresponding to crossover
points of the energy loss processes. We believe that the main source
of the small discrepancies in our calculations was their neglect of
multiple pair-production scattering in a single time step. As a check
on the validity of our mean energy loss approximation in the high
energy region where pion production dominates, we compared our results
to Monte Carlo calculations provided to us by F. Aharonian (using the
code of Aharonian and Cronin\cite{ac}) for monoenergetic injection of
protons at~$10^{21}\,\rm eV$. We obtained good agreement\footnote{The
two calculations differ in the region above~$10^{20.2}\rm\,eV$ in the
following respects: In the Monte Carlo calculation, about 10\% of the
particles above~$10^{20.2}\rm\,eV$ are not scattered and remain
at~$10^{21}\rm\,eV$. In our mean energy loss calculation, these
particles are, of course, distributed continuously, and our spectrum
tracks the Monte Carlo spectrum very closely but differs in overall
normalization by being roughly 25\% higher.} everywhere between the
injection energy and approximately~$10^{20}\,\rm eV$, where pair
production effects and evolution of the CMBR, both neglected in the
code of Aharonian and Cronin, begin to become important.

\section{Contribution to the Cosmic Ray Spectrum of Protons Injected
with Energies Above the Greisen Cut-Off}
\label{HighEnergyResults}

The attenuation length, $L(E)$, for a particle with velocity~$v
\simeq c$ is defined by the equation
\begin{equation}
L = - c E \left(\frac{dE}{dt}\right)^{-1}.
\label{AttLengthEqn}
\end{equation}
The attenuation length for cosmic ray protons in the present universe
as determined from Eq.(\ref{mel}) is plotted in
Figure~\ref{AttLength}. Note that for proton energies below
about~$10^{17.8}\,\rm eV$, cosmological redshift is the only
significant source of energy loss in the present universe.  For
energies between~$10^{17.8}\,\rm eV$ and~$10^{19.6}\,\rm eV$, the
dominant source of energy loss is $e^+e^-$ pair production, whereas
photopion production dominates above~$10^{19.6}\,\rm eV$. By
$10^{19.8} \,{\rm eV}$, the attenuation length is about an order of
magnitude smaller than the Hubble radius, and is dropping rapidly with
increasing energy. Thus, photopion production should significantly
deplete the cosmic ray proton population at energies of $10^{19.8}
\,{\rm eV}$ and higher, although how dramatic a ``cut-off'' one
obtains will depend, to some degree, on the assumed form of the
injection energy spectrum. For definiteness in our terminology, we
will refer to the energy $10^{19.8} \,{\rm eV}$ as the ``Greisen
cut-off'' energy\cite{gre,zk} (This is consistent with the definition
used by Berezinsky and Grigor'eva\cite{bg}). Note that in prior epochs,
the microwave background was at a higher temperature, and the
corresponding ``Greisen cut-off'' energy is thereby redshifted to
roughly the value $(10^{19.8}\,{\rm eV})/(1+z)$, where $z$ denotes the
redshift factor. Since the photon density increases in the past as
$(1+z)^3$ but the Hubble radius decreases only as $(1+z)$, in prior
epochs, the attenuation length for a proton with energy above the
``redshifted Greisen cut-off'' was an even smaller fraction of the
Hubble radius than it is in the present universe.

Despite the prediction of an effective cut-off in the observed cosmic
ray spectrum, cosmic rays have been observed with energies well in
excess of $10^{20} \,{\rm eV}$ \cite{fe2,ake}. One possible
explanation for this fact is that these cosmic rays are not
protons. We shall not consider this possibility in this paper.  If
they are protons, then they must come from a nearby source. It is
possible that we are unusually close to such a source (i.e., the
distance from us to the source is significantly less than the average
distance between sources); in that case, the spectrum of protons we
observe above the Greisen cut-off would not be representative of the
flux occurring in other regions of the universe. Again, we shall not
consider this possibility here, but will assume that the cosmic ray
sources are homogeneously distributed in space and that we are not
unusually close to any single source.  The presence of a cosmic ray
flux above the Greisen cut-off then implies a corresponding injection rate
of above-Greisen-cut-off protons throughout the present universe. If such
protons also were injected at prior epochs in the evolution of the
universe, they will contribute to the presently observed cosmic ray
spectrum at energies below the Greisen cut-off.  The purpose of this
section is to calculate this contribution to the cosmic ray spectrum
under various hypotheses about the strength of the sources in the
past.

If the cosmic ray spectrum above $10^{19.5} \,{\rm eV}$ were
accurately known, the comparison of the observed and predicted spectra
in this energy range (particularly near $10^{19.8} \,{\rm eV}$) would
provide a great deal of quantitative information, which likely would
be sufficient to confirm or rule out models in which the cosmic rays
are protons.  Unfortunately, the data above $10^{19.5} \,{\rm eV}$
suffers from very poor statistics as well as significant random and
systematic uncertainties in energy determinations. For this reason, we
shall not attempt to interpret any nuances in the data in this energy
range reported by the various groups, and merely view the data as
indicating that the differential energy spectrum of cosmic rays above
$10^{19.5} \,{\rm eV}$ appears to be compatible with a $1/E^3$
fall-off, which continues up to and beyond $10^{20} \,{\rm eV}$,
without any dramatic break.

We do not know the initial energy, $E_0$, at which the
above-Greisen-cut-off cosmic rays are injected into the
universe. However, the contribution to the present cosmic ray spectrum
of cosmic ray protons injected at energy $E_0 \gg 10^{19.8} \,{\rm
eV}$ is largely independent of $E_0$.  To see this, consider, first,
the cosmic ray protons which have present energy $10^{19.8} \,{\rm eV}
< E < E_0$. These protons must have been injected within a small
fraction of the Hubble time ago, so the temperature change of the
microwave background can be neglected in calculating their energy
loss. In this energy range, the observed flux between $E$ and $E + dE$
should be proportional to the amount of time the proton spends in this
energy interval, which, in turn, is proportional to~$L(dE/E)$,
where~$L$ was defined in Equation~\ref{AttLengthEqn}. Thus, the
spectral shape of this high energy portion of the present energy
spectrum should be independent of $E_0$---provided, of course, that we
restrict attention to energies $E < E_0$. On the other hand, the
cosmic ray protons which have present energy less than about
$10^{19.5} \,{\rm eV}$ were injected at least~$2\rm\,Gyr$ ago, which
is roughly 50 attenuation times for a proton at~$10^{21}\rm\,eV$ in
the present universe. To a good approximation, the energy loss due to
photopion production for these protons can be treated as causing an
instantaneous decrease of their energy to the redshifted Greisen
cut-off energy, and, thus, the energy they have today will be
essentially independent of $E_0$. Consequently, the predicted spectrum
below about $10^{19.5} \,{\rm eV}$ also does not depend upon $E_0$.
Although the choice of $E_0$ could have a small effect on the present
spectrum between about $10^{19.5} \,{\rm eV}$ and $10^{19.8} \,{\rm
eV}$, this is not significant enough to concern us here.

\begin{figure}
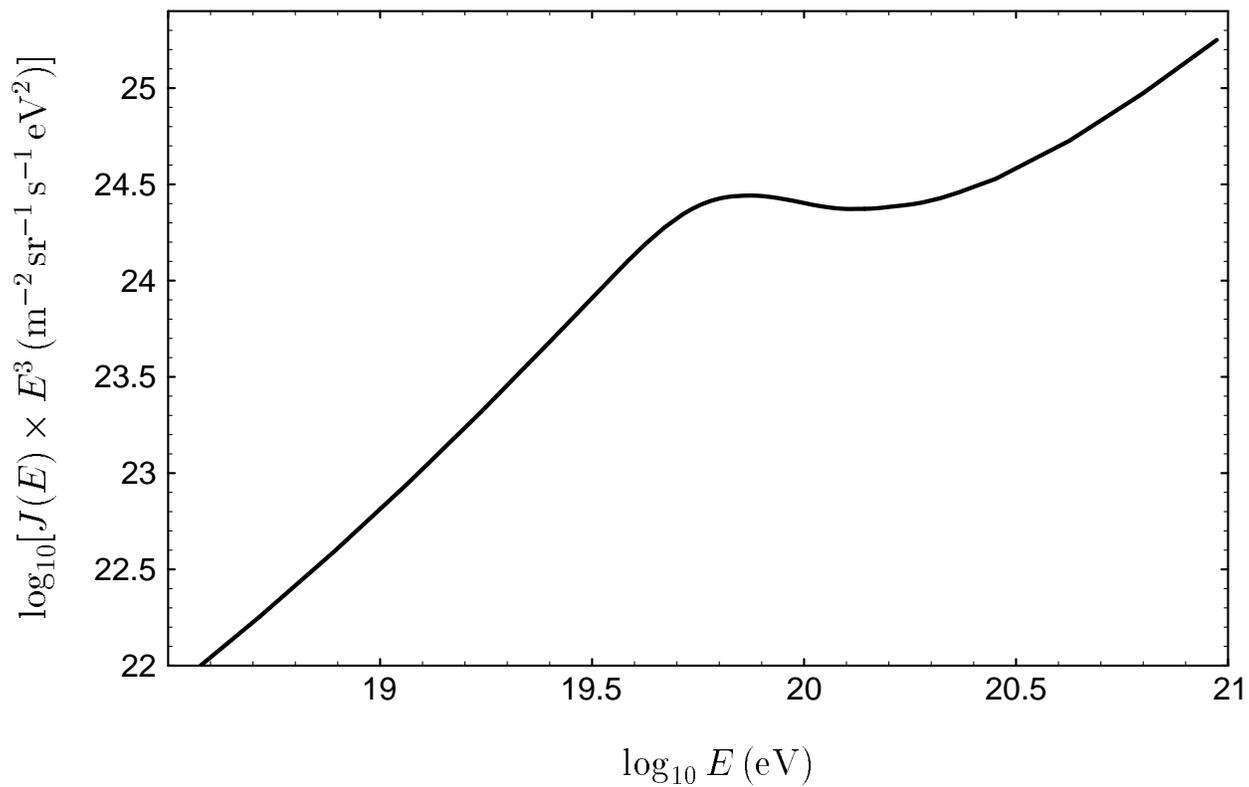

\caption{The differential energy spectrum of cosmic rays resulting
from monoenergetic injection at~$10^{21}\,\rm eV$ with no source
evolution. The overall normalization of this curve was chosen to
correspond to the same present-day injection rate as in
Figure~\ref{HighEnergyWithEvol} below.}
\label{HighEnergyNoEvol}
\end{figure}

Figure~\ref{HighEnergyNoEvol} shows the predicted present energy
spectrum of cosmic rays arising from the injection into the universe
of protons with initial energy $E_0 = 10^{21} \,{\rm eV}$ at a rate
per comoving volume which does not vary with time. Note that the
spectrum shown in Figure~\ref{HighEnergyNoEvol} is the ``hardest''
possible spectrum compatible with the assumptions that the cosmic rays
are protons and their sources are homogeneously distributed in space
and time. By the arguments of the previous paragraph, essentially the
same energy distribution would result from any injection spectrum such
that most of the protons have initial energy much greater than
$10^{19.8} \,{\rm eV}$. (An example of such a differential injection
spectrum is a power law $E^{-\gamma}$ with $\gamma$ near $1$ (or
smaller) and a cut-off (if any) taken to be well above $10^{19.8}
\,{\rm eV}$.) As can be seen from Figure~\ref{HighEnergyNoEvol}, the
predicted spectrum is remarkably compatible with the observed spectrum
at energies above~$10^{19.6} \,{\rm eV}$ (i.e. roughly a~$1/E^3$
spectrum), but lies significantly below a $1/E^3$ spectrum at lower
energies. From these facts, we can immediately draw the following two
key conclusions:
\begin{enumerate}
\item The injection of protons at energies above the Greisen cut-off
at a steady rate compatible with the observed above-Greisen-cut-off
cosmic ray flux could plausibly account for most (or even all) of the
cosmic rays presently observed with energies above $10^{19.6} \,{\rm
eV}$. In any case, there is very little ``room'' for injection of
additional cosmic rays at energies between, say, $10^{19.6} \,{\rm
eV}$ and $10^{20} \,{\rm eV}$.
\item A steady injection of protons at energies
above the Greisen cut-off cannot account for the observed cosmic ray flux
below about $10^{19.6} \,\rm eV$.
\end{enumerate}

The above calculations and conclusions refer to a source of cosmic ray
protons which injects the protons only at energies well above the
Greisen cut-off. However, it would seem more reasonable to assume that
a realistic source would inject protons with a distribution of
energies that extends to below the Greisen cut-off energy.  If so, one
may ask whether such a source could plausibly account for cosmic rays
with energy below $10^{19.6} \,{\rm eV}$ as well.  However, this does
not appear to be at all likely: In order to significantly improve the
agreement between predicted and observed spectra at energies below
$10^{19.6} \,{\rm eV}$, one would need a differential injection
spectrum at these energies that is ``softer'' than $E^{-2}$. On the
other hand, in order to maintain the absence of a dramatic drop in the
predicted flux above the Greisen cut-off energy, one needs an
injection spectrum considerably ``harder'' than $E^{-2}$ at energies
above $10^{19.8} \,{\rm eV}$. Thus, unless nature has contrived to put
in a break in the injection spectrum very close to the Greisen
cut-off, one cannot simultaneously account for the observed flux below
$10^{19.6} \,{\rm eV}$ and the absence of a dramatic drop in the
predicted flux above $10^{19.8} \,{\rm eV}$.

In the next section, we will argue against the possibility that the
cosmic ray flux between $10^{18.5} \,{\rm eV}$ and $10^{19.6} \,{\rm
eV}$ can be understood as a continuation of the $1/E^3$ spectrum
observed at lower energies.  For the remainder of this section, we
shall investigate the extent to which the presently observed cosmic
ray flux below $10^{19.6} \,{\rm eV}$ could be explained by injection
of protons at initial energy above the Greisen cut-off by sources
which were more numerous in the past. We will model the comoving
source density by an exponential time dependence, i.e., we will take
the injection rate of above-Greisen-cut-off protons to vary with time
as $h(t) = \exp(-t/\tau)$, as would be appropriate if the source were
an unstable, decaying particle produced in the big bang; exponential
time dependence is also used to model quasar
activity\cite{sgquasar}. However, we would expect the results for a
power law time dependence (presumably appropriate for the decay of
particles produced by cosmic strings) to be qualitatively similar
(except for effects occurring in the very early universe if one
extrapolates the power law dependence all the way back to the big
bang).

\begin{figure}
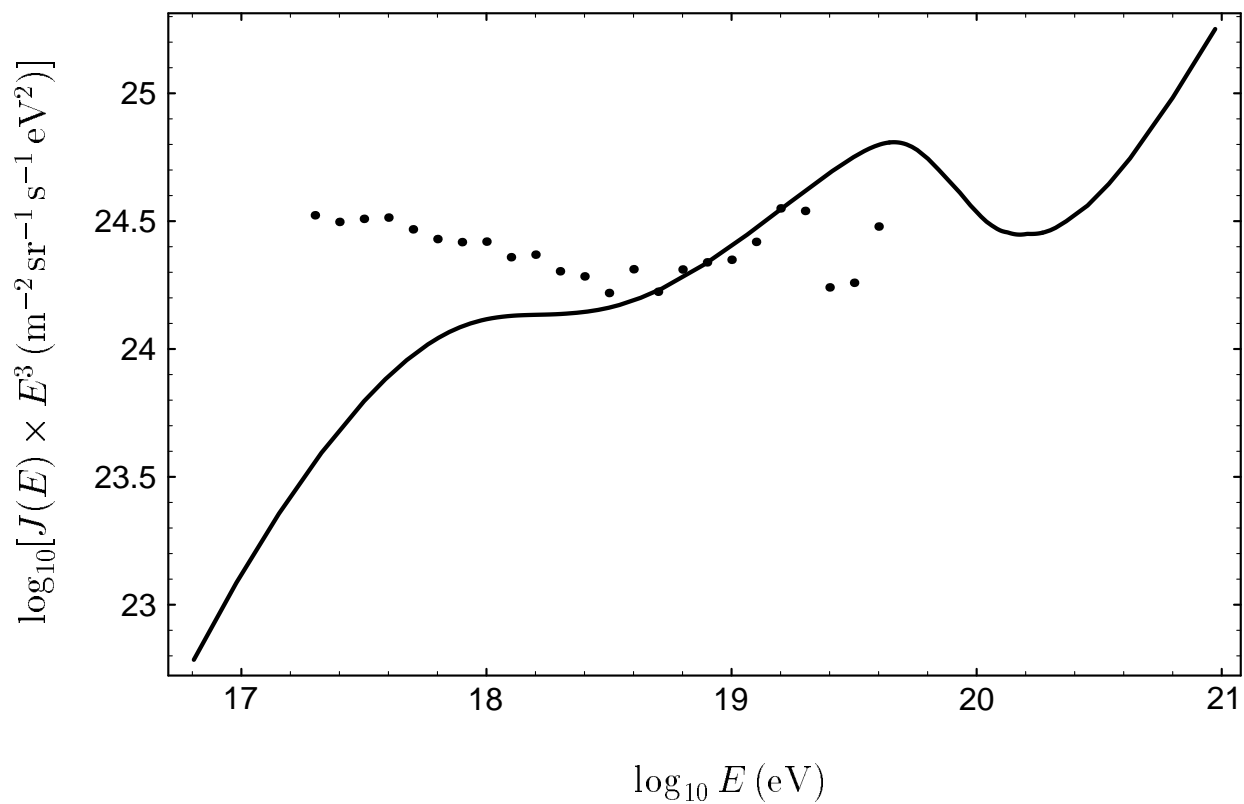

\caption{The differential energy spectrum of cosmic rays due to
monoenergetic injection of protons at~$10^{21}\,\rm eV$ with source
evolution given by~$\exp(-t/\tau)$ with $\tau=t_0/12$ (solid line) as
compared to the spectrum reported by the Fly's Eye group (points). The
normalization of the curve corresponds to a present day injection rate
of~$5\times10^{-46}\,{\rm cm}^{-3}\,{\rm s}^{-1}$.}
\label{HighEnergyWithEvol}
\end{figure}

Figure~\ref{HighEnergyWithEvol} shows the results of our attempt to
fit the Fly's Eye stereo energy spectrum data\cite{fe1} down to as low
an energy as possible. Note that the fit is very poor for the three
highest energy points between~$10^{19.4}\rm\,eV$
and~$10^{19.6}\rm\,eV$, but the statistics in this energy region are
very poor (two or three events per bin) and this ``dip'' is not
observed by other groups (see, e.g., the graphs in \cite{fe1}). The
optimal choice of~$\tau$ appears to be~$\tau \simeq t_0/12 =
1.25\rm\,Gyr$.  Smaller values of~$\tau$ (stronger evolutions) would
produce a larger ``bump'' at about~$10^{19.8}\,\rm eV$ due to enhanced
injection of particles at earlier times. Larger values of~$\tau$
(weaker evolutions) tend to produce spectra which more closely
approximate a ``flat''~$1/E^3$ spectrum above~$10^{19.6}\,\rm eV$, but
fail to fit the data in the lower energy regions, falling off
dramatically somewhere between~$10^{18.5}\,\rm eV$ and~$10^{19.6}\,\rm
eV$. In any case, even with strong evolution it does not appear
possible to fit the data below the bottom of the Fly's Eye ``dip''
at~$10^{18.5}\,\rm eV$. Finally, note that our predicted spectrum has
a minimum (when plotted in this manner) slightly
above~$10^{20}\rm\,eV$, which could be compatible with the existence
of a ``gap'' in the cosmic ray spectrum hinted at by the observational
data.

Our calculations show that the present cosmic ray proton energy
of~$10^{18.5}\rm\,eV$ corresponds to an above-Greisen-cutoff injection
at redshift~$z \simeq 1/2$. (Even for a Hubble constant of
$H_0=75\,{\rm km}\,{\rm s}^{-1}\,{\rm Mpc}^{-1}$ the injection would
have occurred at~$z \simeq .6$.) Note that it follows that in any
model where the cosmic rays are protons, all cosmic rays with present
energy~$\geq 10^{18.5}\rm\,eV$ must have been injected quite recently
(namely at~$z \leq 1/2$). In the case of the model above, the optimal
choice of~$\tau$ corresponds to an injection rate at~$z=1/2$ which
is~$\sim 200$ times greater than the present injection rate. This is
much too rapid a change to plausibly result from the evolution of any
ordinary astrophysical source.

However, one possible candidate for the source of ultra-high-energy
cosmic rays is an unstable relic particle. As will be seen below, in
order to avoid making an unacceptably large contribution to
the~$\gamma$-ray background, it will be necessary for this particle to
decay efficiently into protons. Hence, we will assume that this
particle is a baryon and produces one proton per decay. Since the
particle has a comoving number density,~$n_{\rm X}$, which varies with
time as $n_{\rm X} = n_0 \exp(-t/\tau)$, the proton injection rate per
unit volume today,~$r_0$, is given by
\begin{equation}
r_0 = -\left.\frac{dn_{\rm X}}{dt}\right|_{t_0}
	=\frac{n_0}{\tau} e^{-t_0/\tau}.
\end{equation}
Using the values $r_0 = 5\times 10^{-46}\rm\,cm^{-3}$ and $\tau =
t_0/12 = 4\times 10^{16}\rm\,s$ which were used to fit the data in
Figure~\ref{HighEnergyWithEvol}, we obtain
\begin{equation}
n_0 = 3\times 10^{-24}\rm\,cm^{-3}.
\end{equation}
This corresponds to a present number density of~$2\times
10^{-29}\rm\,cm^{-3}$.

The highest energy cosmic ray ever observed had an energy of~$3\times
10^{20}\rm\,eV$\cite{fe2}. This would imply a lower limit on the mass
of the decaying particle of order $m_{\rm X} =
10^{21}\rm\,eV$. Choosing this value of $m_{\rm X}$, we find that in
the present universe $m_{\rm X} n_{\rm X} = 2\times
10^{-8}\rm\,eV\,cm^{-3}$, which is roughly $10^{-11}$ of the critical
density. Thus, such a hypothetical particle would make a negligible
contribution to the mass density of the present universe. Even at time
$t\sim\tau$ (when a large fraction of the particles had not yet
decayed), the contribution to the mass density of the universe would be
only of order $10^{-6}$ of the critical density, which is to small to
have an influence on the dynamics of the universe, though possibly
large enough to produce astrophysically interesting effects.

The existence of such a decaying particle would produce some
potentially observable consequences, and, hence, observation may be
used to set bounds on its mass, lifetime, and abundance. The most
relevant bound arises from a consideration of the effect of the
decaying particle on the $\gamma$-ray background. The high-energy
cosmic rays produced by the decay of the relic particle interact with
the thermal background radiation, producing $e^+ e^-$ pairs. These
pairs subsequently inverse-Compton scatter on photons in the
background radiation, producing more high-energy photons, many of
which themselves then pair-produce on the thermal background. This
electromagnetic cascade continues until the highest energy photons
drop below the threshold for pair-production. A well-defined
prediction is thus obtained for the~$\gamma$-ray spectrum resulting
from this cascade, and a bound on the mass and abundance of the
decaying particle (which depends only weakly on its lifetime for the
range of lifetimes relevant here) results from the requirement that
the flux of $\gamma$-rays from the cascade not exceed the observed
flux. The bound derived by Ellis~et~al.\cite{ellis} arises from
observation of the gamma ray flux at $170\rm\,MeV$ and yields the
limit\footnote{It should be possible to strengthen this bound (by
roughly a factor of 5) using the more recent data analyzed
in~\cite{osb}, since this data extends to higher $\gamma$-ray energies
(though with larger uncertainties). However, it undoubtedly will be
necessary to undertake a much more complete and accurate analysis of
the predicted spectrum in order to set any firm bounds.} $m_{\rm X}
n_0 < 4\times 10^{-3}\rm\,eV\,cm^{-3}$. This is remarkably close to
the value of $m_{\rm X} n_0 = 3\times 10^{-3}\rm\,eV\,cm^{-3}$
obtained using the mass and abundance given above. However, it should
be noted that our calculated value of~$n_0$ is very sensitive to the
assumed value of~$\tau$. A lifetime of~$t_0/10$ rather than~$t_0/12$
would still provide an acceptable fit to the Fly's Eye Data, but would
result in a value of the number density an order of magnitude below
the observational limit.  Thus it appears that the above relic
particle decay model is compatible with observational constraints.
However, these constraints would appear to require the particle to be
a baryon of mass not much larger (or smaller) than~$10^{21}\rm\,eV$,
which decays in a ``clean" manner.

\section{Contribution to the cosmic ray spectrum of protons injected
with a~$1/E^3$ spectrum}
\label{LowEnergyResults}

We have shown in the previous section that injection of protons above
the Greisen cut-off cannot account for the observed cosmic ray
spectrum below about~$10^{18.5}\,\rm eV$ even with strong source
evolution. The observed spectrum has a~$1/E^3$ energy dependence in
the region below~$10^{17.6}\,\rm eV$ extending down to~$10^{15}\,\rm
eV$. In this energy regime, cosmological redshift dominates and
therefore the power index of the injection spectrum should be
preserved. Hence, it is natural to ask to what extent the observed
spectrum can be explained by a~$1/E^3$ injection spectrum.

The statistics of the available data are much better in this low
energy region, and the following details of the observed spectrum are
worth noting.  The differential spectrum reported by the Fly's Eye
group\cite{fe1} has a~$1/E^3$ dependence up to
approximately~$10^{17.6}\,\rm eV$, but it then steepens in the region
between~$10^{17.6}\,\rm eV$ and~$10^{18.5}\,\rm eV$ (the Fly's Eye
group reports a power index of $-3.27$) and then flattens
between~$10^{18.5}\,\rm eV$ and~$10^{19.6}\,\rm eV$ (Fly's Eye reports
a power index of $-2.75$). When plotted in the usual way ($J[E]\times
E^3$) this results in the so-called ``dip'' in the differential
spectrum with its minimum at~$10^{18.5}\,\rm eV$.

\begin{figure}
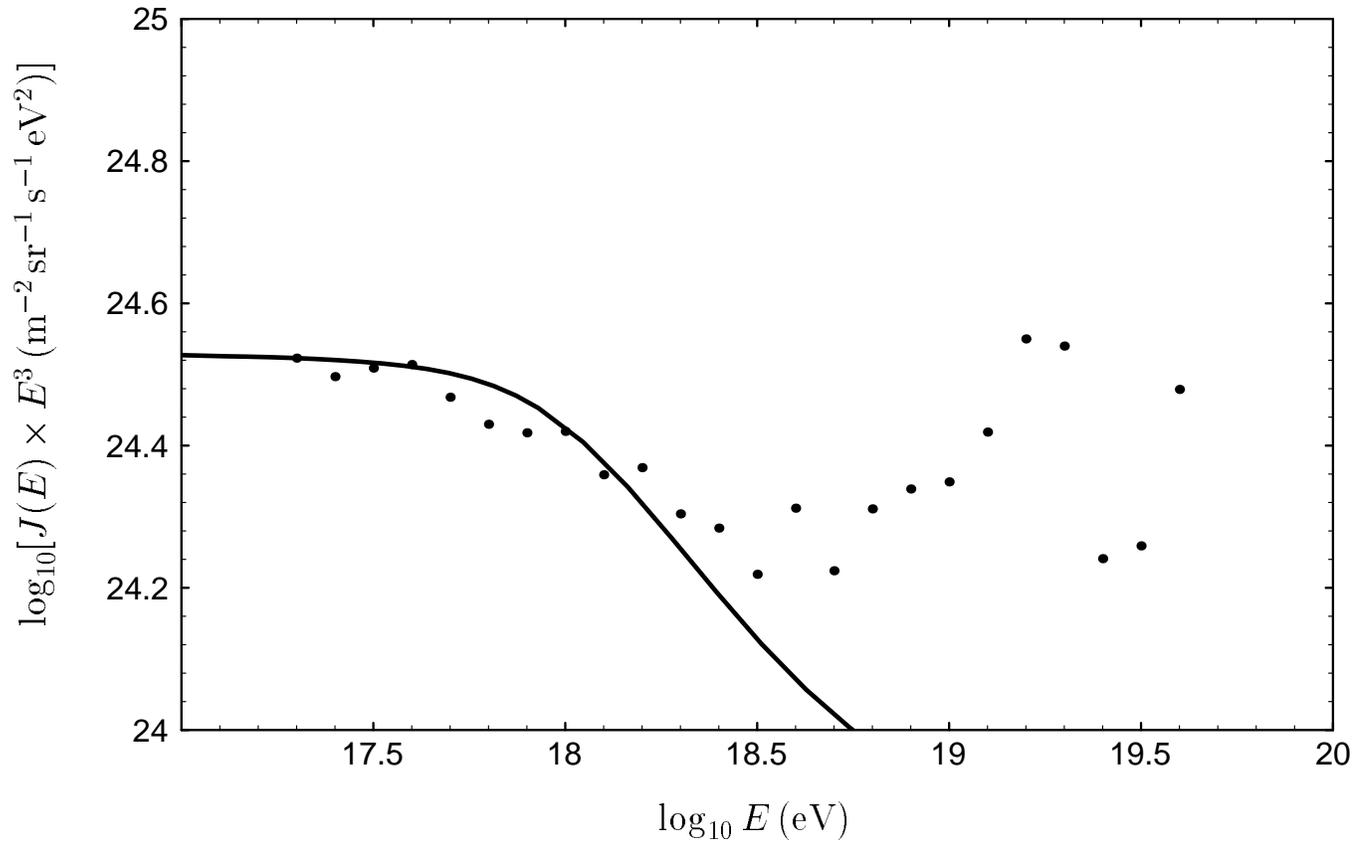

\caption{The differential spectrum of cosmic rays due to a~$1/E^3$
injection spectrum of protons with no source evolution (solid line) as
compared to the spectrum reported by the Fly's Eye group (points).}
\label{LowEnergyNoEvol}
\end{figure}

In Figure~\ref{LowEnergyNoEvol} we plot the spectrum which would
result from protons being injected with a~$1/E^3$ spectrum and no
source evolution, normalized to the Fly's Eye data
at~$10^{17.3}\rm\,eV$.  It should be noted that, apart from this
overall normalization, there are no free parameters in our plot
(although small adjustments to the curve could be produced by varying
the Hubble constant or considering evolutionary effects [see
below]). The agreement with the Fly's Eye data up to
about~$10^{18.5}\rm\,eV$ is very good, but diverges strongly from the
data above~$10^{18.5}\rm\,eV$.  Two conclusions can be drawn from this
result.  First, the steepening of the observed spectrum is very
naturally explained by the depletion of the region
above~$10^{17.6}\,\rm eV$ by pair production effects. This can be
viewed as supporting our hypothesis that the cosmic rays in this
energy regime are protons of extragalactic origin. Second, this
depletion produces a steepening of the spectrum which persists well
beyond the bottom of the ``dip'' at~$10^{18.5}\,\rm eV$, producing the
strong disagreement with the data above this energy noted above. This
depletion is unavoidable under the hypothesis that the cosmic rays are
protons of extragalactic origin, so, under this hypothesis, a~$1/E^3$
(or softer) injection spectrum spectrum with no source evolution
cannot explain the observed cosmic ray flux above~$10^{18.5}\rm\,eV$.

\begin{figure}
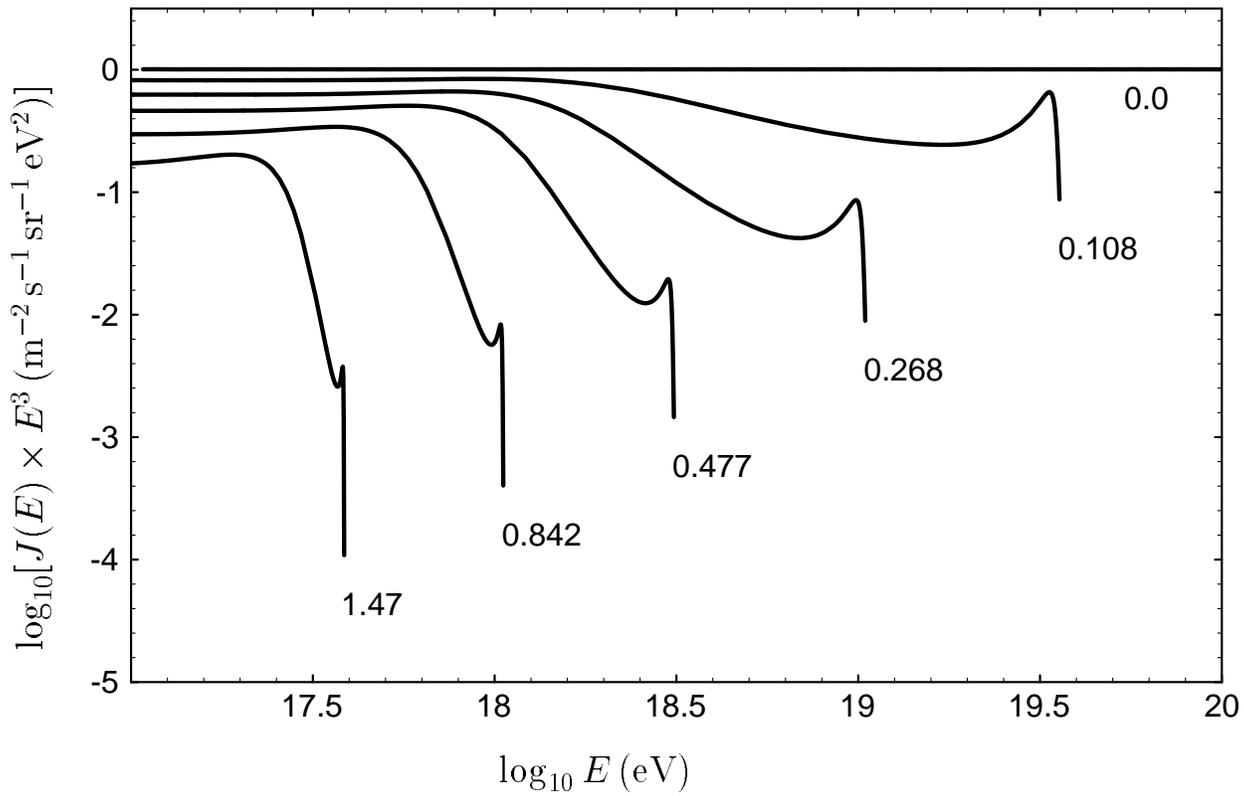

\caption{The relative contributions to the predicted spectrum from
several epochs under the assumption of a~$1/E^3$ injection spectrum
and no source evolution. (The overall normalization is chosen in order
to make the contribution from the present universe equal to
unity.) The labels on the curve give the redshift value of the epoch.}
\label{LowEnergybyEra}
\end{figure}

In order to show the possible effects of source evolution on this
spectrum, we plot in Figure~\ref{LowEnergybyEra} the relative
contributions of several epochs spaced uniformly in time. The
integration which produces the predicted spectrum today in the case of
no source evolution is equivalent to simply summing these curves.
(However, in our calculations, we used a much finer time spacing.)
Therefore, by weighting these curves appropriately, one can predict
the shape of the spectrum for any source evolution model. It can be
seen from Figure~\ref{LowEnergybyEra} that unless the source evolution
is extremely strong the contributions from redshifts~$z>1$ will be
insignificant at energies above~$10^{17.8}\rm\,eV$. It is also clear
from Fig.~\ref{LowEnergybyEra} that any source evolution function
which is stronger in the past will simply produce a steepening which
is sharper and which begins at a lower energy than is seen in the case
of no evolution above. Finally it can be seen (as we have verified by
detailed calculations) that no choice of evolutionary model can
provide a good fit to the Fly's Eye data at energies
above~$10^{18.5}\,\rm eV$.

\section{Conclusions}
\label{Conclusions}

We have shown that, under our hypotheses that the cosmic rays are
protons of extragalactic origin and that their sources are
homogeneously distributed in space, a~$1/E^3$ injection spectrum
plausibly accounts for the observed cosmic ray spectrum up
to~$10^{18.5}\,\rm eV$ but cannot account for the observed flux at
higher energies. We have also shown, under the same hypotheses, that
the injection of protons with energy above the Greisen cut-off at the
rate needed to account for the presently observed above-Greisen-cutoff
flux could plausibly account for the observed spectrum down to an
energy of~$10^{19.6}\rm\,eV$. However, we have demonstrated that, in
order to explain the observed flux between~$10^{18.5}\,\rm eV$
and~$10^{19.6}\,\rm eV$ under our hypotheses, one must postulate
either that (A) the above-Greisen-cutoff source was very much stronger
in the past, (B) that the injection spectrum of the low energy source
becomes considerably harder than $1/E^3$ above $10^{18.5} \,{\rm eV}$,
or (C) that there is yet another source of cosmic rays injecting
particles between $10^{18.5} \,{\rm eV}$ and $10^{19.6} \,{\rm
eV}$. The decay of an ultra-massive particle with a lifetime of
order~$10^9\rm\,yr$ would provide an apparently viable mechanism for
possibility (A). However, we leave it to the reader's judgement to
determine if this fact should be viewed as evidence in favor of the
existence of such a particle or as a demonstration of the lengths to
which one must go in order to account for the observed high energy
cosmic ray spectrum under our above assumptions.

Unless there is a major theoretical breakthrough in our understanding
of the physical mechanisms underlying the sources of cosmic rays it is
likely that further advances in our understanding of ultra-high-energy
cosmic rays will require better observational data. In particular, a
much more precise determination of the energy spectrum in the vicinity
of~$10^{19.8}\rm\,eV$ should provide a stringent test of any model in
which the cosmic rays are protons.

\section*{Acknowledgements}
We wish to thank F. Aharonian, J. Cronin, R. Ong, and G. Sigl for
helpful discussions and F. Aharonian, G. Sigl, and P. Sommers for
reading the manuscript. Special thanks go to F. Aharonian for
providing us with the results of a Monte Carlo calculation. This
research was supported in part by National Science Foundation grant
PHY-9220644 to the University of Chicago.

\end{document}